# Video Killed the Writing Assignment

Nicolle Zellner[1,*]

[1,*] Department of Physics, Albion College, Albion, MI, USA 49224, nzellner@albion.edu

* To whom correspondence should be addressed

**Abstract**
An introductory Astronomy survey course is often taken to satisfy a college graduation requirement for non-science majors at colleges around the United States. In this course, material that can be broadly categorized into topics related to "the sky", "the Solar System", "the Galaxy", and "cosmology" is discussed. Even with the wide variety of topics in these categories, though, students may not be 100% interested in the course content, and it is almost certain that a specific topic about which a student wishes to learn is not covered. To at least partly address these issues, to appeal to all of the students in this class, and to allow students to explore topics of their choice, a video project has been assigned to students at Albion College as a class activity. In this assignment, students are asked to create a video of a famous (or not) astronomer, astronomical object or discovery, or telescope observatory to present to the class. Students work in pairs to create a video that is original and imaginative and includes accurate scientific content. For this project, then, students use a familiar technology and exercise their creativity while learning a little (or a lot of) science along the way. Herein data on types and topics of videos, examples of videos, assignment requirements and grading rubrics, lessons learned, and student comments will be discussed and shared.

## INTRODUCTION

Introductory Astronomy courses are a mainstay of course offerings in colleges around the United States and Canada (e.g., Fraknoi 1998), reaching almost 200,000 American students at four-year colleges and over 50,000 American students at two-year colleges in 2011-2012 (Mulvy and Nicholson, 2014). For generations, the night sky has spurred curious eyes to look up, and telescopes such as the Hubble Space Telescope and Cassini have sent back to Earth amazing images of celestial bodies that are appreciated by scientists and non-scientists alike. Moreover, the origin and fate of the Universe have stimulated scientific and philosophical debate from 1921 (e.g., Hoskin 1976; Curtis 1921; Shapley 1921) to the present (e.g., Maeder 2017; Dam et al. 2017). Given the popularity of astronomical topics in general (e.g., Guglielmi 2018; Redi et al. 2018), including the thought-provoking possibility of habitable environments on several of the ~5300 exoplanets and exoplanet candidates (http://exoplanets.org/, accessed 5/15/18) observed by the Kepler spacecraft, it is no wonder that "[e]nrollments in introductory astronomy courses have been steadily increasing over the years and are now about 10% higher than what they were a decade earlier" (Mulvey and Nicholson, 2014). Importantly, introductory courses like Astronomy may be responsible for increased scientific literacy in the United States (Wittman 2009; Hobson 2008).

While there is no set selection of topics, it is typical for "the universe" to be taught in a one-semester survey course (e.g., Slater et al. 2001). Thus, in the Introductory Astronomy course (Phys 105) for non-science majors at Albion College, it is usual to cover 13.7 billion years of Astronomy in 15 weeks (i.e., one semester) of classes. The list of topics includes sky motions and constellations, the Solar System, Astrobiology, and galaxies and cosmology. Additionally, this survey Astronomy course may be the only "mathy" course (e.g., Fowler 2011) a student takes at Albion College, and so Kepler's Laws, the Leavitt Law, and Hubble's Law are also included. Finally, in keeping with trends observed by Slater et al. (2001), activities in the classroom and in the laboratory setting are designed to inspire life-long learning. Thus, many of the "Astro 101 goals" (Partridge and Greenstein, 2004) are met. However, given the semester constraint, much material (or detail in the topics presented) is omitted.

The format of a course such as this one varies from institution to institution and also from instructor to instructor at these institutions, though oftentimes a semester-long project or term paper is assigned so that students can investigate in-depth a topic of their choice. As the only Astronomy instructor at Albion College, I have a lot of freedom in how I teach my class, and after several semesters of assigning such a term project, it became apparent that students did a superficial study of the topic and did not learn or recall anything about that topic, as reported anecdotally by students. As a result, in Fall 2008, a video assignment, in which students create a video on a topic of their choice, replaced the term paper. In order to incentivize the students to take time to create a quality video, the project counts as 10% of their overall grade and one (or more) class period(s) is(are) spent watching the videos together as a class. The students also vote to determine the "People's Choice Award" winning video, whose creators receive a special prize from me.

Overall, the response to this project has been very positive and students look forward to this assignment. From the perspective of the instructor, students get to implement a technology they are already using to be creative and imaginative, and learn a little (or a lot of) science along the way. And everyone has fun watching the videos together as a class. Since Fall 2008, 140 videos have been created, involving 356 Albion College students, 75% of whom were registered for the class.

## INVESTIGATION
### Assignment
Typically, college students write a lot of papers, and based on conversations with students at Albion College, it appears that they start, review, and finish the final paper within 24-48 hours of its due date. This timeline can be stressful for the student and the final result can be exasperating for the instructor, as punctuation, spelling, and grammar rules are usually ignored (e.g., Nunberg 1983) and plagiarism can be rampant (e.g. Ali et al. 2012). The video assignment aims to alleviate at least some of these concerns.

Additionally, students work together and interact with each other (Prather et al. 2009a,b) to investigate a topic of their choice (e.g., Slater 2015) and to develop collaborative skills that will carry beyond the classroom (Partridge and Greenstein, 2004; Slater et al. 2001). The goal of the project is for the students to be creative yet still include accurate scientific content in their video. The instructions for this assignment can be found in Appendix 1.

The video can focus on almost any topic relevant to Astronomy, including documenting how a great discovery was made; chronicling the life of an astronomer; recording the moment something was discovered at an observatory; or anything else that is appropriate and subject to final approval. As a result of choosing the topic they find interesting, students take ownership of their video; topics thus range from women in astronomy (10 videos) to dark matter (2 videos). Topics can be broadly categorized, with biographies of astronomers and scientists being most popular, followed by Solar System topics and black holes (Figure 1, Table 1). In order to avoid duplicates in any one class, topics are registered as "first come, first served", and students are encouraged to register topics with me as soon as possible.

The style of the video is also left to the imagination of the students. However, in 140 videos, students have opted for just a few different ways in which to film their video. These include documentary (45 videos), in which the students describe a topic with live characters, action clips, and pictures; voice over (22 videos), a type of documentary that includes only pictures or video clips (and no people), with a student verbally describing what is being shown; spoofs of TV shows (20), in which the students remake a known TV show (e.g., sit-com, game show); and reenactment (17 videos), in which students act out a skit. However, there have also been 12 interviews of the "famous" person, three raps, and multiple individual presentations (Figure 2).

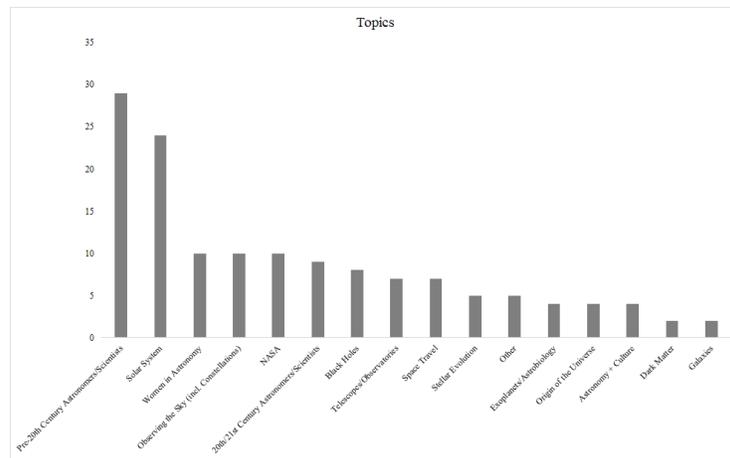

**Figure 1.** List and number of topics presented in 140 videos at Albion College in 18 semesters.

**Examples**

Early on, videos leaned heavily toward historical events and people (Goal #8 in Partridge and Greenstein, 2004) but in the past few semesters, they have focused more on space missions and women and/or underrepresented scientists or astronauts, perhaps reflecting news-making events in the nation. People in general are the subject of most of the 140 videos, while black holes and Solar System objects are the subject of choice for students who want to create a video about a celestial object (Figure 1).

A few specific examples of videos include an interview with a galaxy that was eating another galaxy (i.e., galactic cannibalism), a sit-com centered on Copernicus and his quirky friends (in the style of "Seinfeld"), an episode of "The Office" that portrayed the discovery of the cosmic microwave background radiation by Penzias and Wilson, and "Mythbusters" busting the Moon-landing hoax. Students have re-enacted the trial of Galileo, the formation of a nebula by "cooking one up" in the kitchen (and sharing the cake during the video viewing!), and the discovery of Neptune with a colorful cast of "astronomers" from four different countries. Finally, students in the class have recruited friends to form teams that compete on their own "Family Feud" and "Survivor: Astronomy" game shows that focused on knowledge of Solar System facts. Some students, however, exclude any real people or images in their videos, instead choosing to use their creative talents in other ways (Figure 3).

The "People's Choice Award" winning videos are showcased on my personal blog at campus.albion.edu/nzellner and other videos can be found by searching on "Albion College Astronomy" on youtube.com.

**Grading**

Broadly speaking, the video assignment falls into the category of "digital storytelling" because students have the option to combine photographs, video, animation, sound, music, and text in order to share a story. In this class, students are provided with a list of requirements and the grading rubric ahead of time (Stevens and Levi, 2013). In particular, the video needs to include a title and a reference list, the latter of which is also turned in as a hard (or electronic) copy. Videos need to be posted on youtube.com or on our college's public drive for easy access and viewing. Most recently, students have also been posting videos on a shared Google drive. The length of videos has ranged from 5 minutes to 10 minutes, depending on the size of the class, but most assignments ask for a video of length 5:30 ± 30 seconds. This appears to be the optimal length, both in terms of production and viewing. Student are required to email the link and title to me no later than 11:59 pm of the day before class viewing so that the videos can be easily and quickly uploaded at the start of class, thereby saving time. Student are also asked to test the video on the equipment in the classroom and are required to be in attendance for all days of viewing. Finally, the students are instructed to pay attention to content, scientific accuracy, and creativity (as described in the rubric), and to have fun!

Student videos are evaluated by both their classmates and by me, according to the criteria listed in Table 2, though many other evaluation criteria could be used (Schrock 2018). In general, the best quality videos received the highest scores, regardless of the academic caliber of the student(s) who created the video (in contrast to results of the Sadler and Good (2006) study). Each criterion here is intentionally broad, allowing for the range of topics and styles selected by the students. The final grade is determined by an average of classmates' evaluations (33%) and by my evaluation (67%) and counts for 10% of their overall grade (i.e., the equivalent of one exam). There is a -10% penalty added to an individual's video grade for not adhering to the requirements, including not participating in evaluating classmates' videos, failure to show up to watch the videos, and/or delay of viewing. As previously mentioned, the students who create the video judged best by the class receive a special prize, usually some type of NASA swag or other astronomy-themed item.

**Table 1.** Examples of topics included in 140 videos in 18 semesters.

| Topic | Number |
|---|---|
| **Women in Astronomy** | **10** |
| **20th/21st Century Astronomers/Scientists** | **9** |
| **Black Holes** | **8** |
| **Solar System** | **24** |
| Jovian Planets | 7 |
| Pluto | 6 |
| General | 4 |
| Mars | 3 |
| Comets/KBOs/NEOs | 3 |
| Titan | 1 |
| **Pre-20th Century Astronomers/Scientists** | **29** |
| Galileo | 6 |
| Tycho Brahe | 5 |
| Giordano Bruno | 5 |
| Copernicus | 4 |
| Kepler | 2 |
| Messier | 2 |
| Isaac Newton | 2 |
| The Herschels | 2 |
| Guiseppi Piazzi | 1 |
| **Telescopes/Observatories** | **7** |
| **Observing the Sky (incl. Constellations)** | **10** |
| **Exoplanets/Astrobiology** | **4** |
| **Origin of the Universe** | **4** |
| **Space Travel** | **7** |
| **NASA** | **10** |
| Missions | 6 |
| Astronauts | 4 |
| **Stellar Evolution** | **5** |
| **Astronomy + Culture** | **4** |
| **Dark Matter** | **2** |
| **Galaxies** | **2** |
| **Other** | **5** |

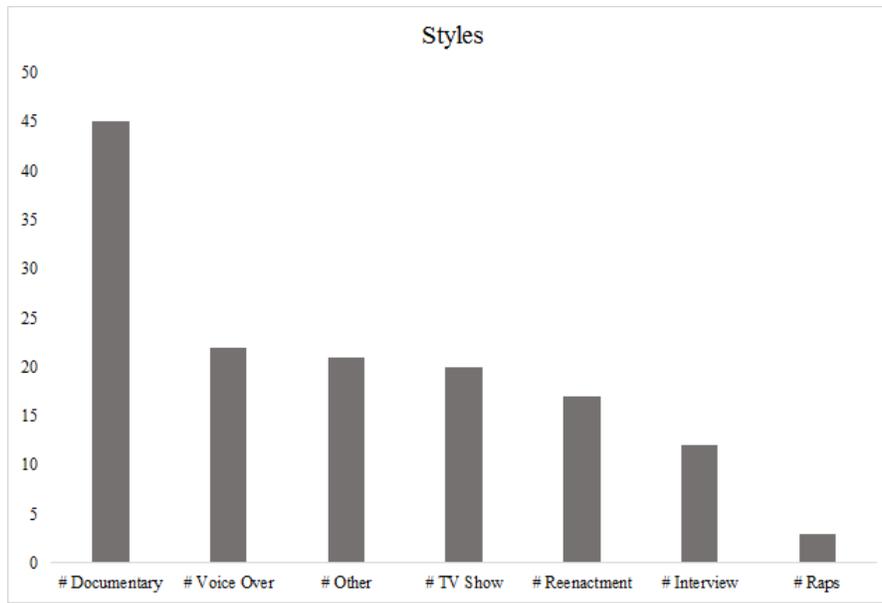

**Figure 2.** List and number of styles in which the 140 videos at Albion College were presented.

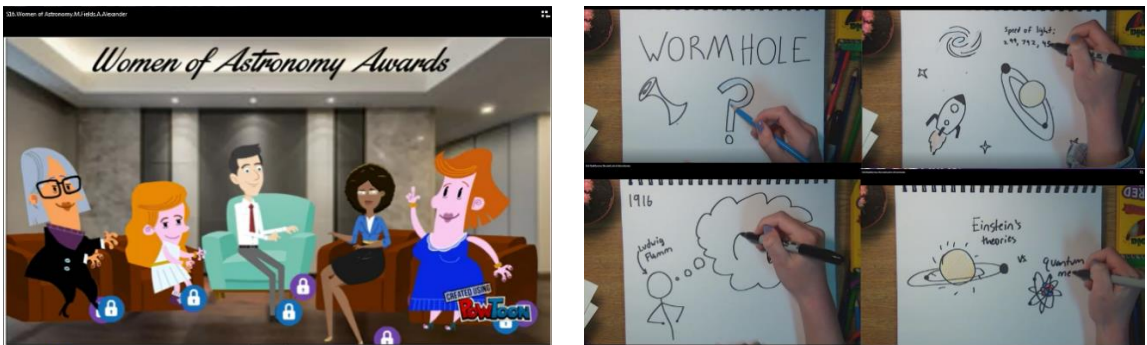

**Figure 3.** Snapshots from two videos where students illustrated the topic of their video, rather than re-enacting or dramatizing it. Video image credits: A. Alexander and M. Fields (L) and N. Moreno and N. Kidman (R).

**Table 2.** Grading criteria as listed on the assignment rubric.

| Grading Criterion | Explanation | Grader |
|---|---|---|
| Content | appropriate organization<br><br>relevant images, animations, or other illustrations<br><br>adequate audio | classmates<br>instructor |
| Science | understandable<br><br>appropriate level | classmates<br>instructor |
| Delivery | appealing<br><br>entertaining | classmates<br>instructor |
| Scientific Accuracy | | instructor |
| Adherence to Requirements | | instructor |

**STUDENT FEEDBACK**

Students are (usually) excited to create a product that is shared with their classmates, which typically does not happen in the case of research term papers. Thus, the overall opinion of the students is that this is an interesting class project that allows them to be creative, share that creativity with their classmates, learn some science, and also develop video film and editing skills that carry beyond the Astronomy classroom.

Comments from students about the project include "procrastination is not a friend". This is because creating a script, filming, and editing can take up a lot of time, especially for the novice. This, however, is becoming rarer as personal devices become more commonplace as student possessions and as students are becoming more adept at using these devices as video cameras. Once the video project is assigned students have about six weeks until the video is due. I have learned to remind the students almost weekly about the project. Importantly, I have now instituted a deadline for the topic (~1.5 weeks after the project is assigned) and announce milestones up until the due date. For example, with about three weeks until the due date, I remind students that they should have begun researching their topic and writing a script. Recent feedback from students indicates that they spend anywhere from two to 10 hours on researching the topic and producing the video. Only three times in 18 semesters have videos been turned in late or not at all.

Another common student comment is that "solo is no good". As a result of the several different tasks related to creating a video, pairs of students are able to "divide and conquer" them and in the end, produce much higher quality videos than students who work by themselves, with a few exceptions. Over the 18 semesters during which this project was assigned, teams of females and small teams of male/female worked well together, consistent with findings of Adams et al. (2001) who found in one case study that "females were categorized as watching passively and or disengaged significantly more frequently when working in groups that contained uneven numbers of males and females." Therefore, keeping the groups small (i.e., in pairs) is ideal and optimizes both video quality and student participation in the project.

As previously indicated, the overall response to the video project is positive. Recent student comments include:

- "The video project was actually really fun. It was a lot of work, but I enjoyed it."
- "It was a fun and creative way to learn about new topics that I wouldn't cover on my own time."
- "It was a fun take on a science project."
- "I enjoyed that I could take it in any direction, so I learned about the scientific aspects of topics in religious studies I enjoy."
- "If people took the project seriously, I learned a lot from their videos and enjoyed these days in class".
- "I could not believe the lack of effort… just read from a Wikipedia page on an astronomer…so boring I couldn't absorb anything."
- "The video project is a long process but overall a blast!"
- "I enjoyed this project because I learned a lot without having to sit through a lecture."
- "I liked learning…something we have not discussed or learned about."
- "Overall this was a fun experience."
- "I enjoyed being able to choose a topic of interest and explore it in greater detail."

**LESSONS LEARNED**

From the instructor perspective, aside from student reminders and one calculation to determine the final grade, the student video project does not require a lot of extra time or oversight, and the reward is a variety of (mostly) entertaining videos! Semester to semester I am consistently impressed by the quality and creativity of (most of) the videos produced by the students. Rarely did I have to answer questions on how to actually make the video, directing them to our campus Instructional Technology department for any question(s) I could not answer; staff there were more than happy to help students with the equipment. Additionally, aside from using up to two class periods to view the videos together, the assignment is also not as time-consuming as grading term papers. Issues that did arise included videos that were too long, inappropriate content and/or sources, too-small font, poor sound quality, or students who were not in attendance during class viewing.

## Quantity and Quality

Most semesters, instructions indicated that the length of the video should be 5:30 ± 30 seconds and that anything over 6 minutes would not be considered in viewing or grading. However, I gave the students a little bit of lee-way (usually ±45 seconds), only reducing their grade substantially if they were very much over the time limit or very much under it. For example, one student group turned in a video that was twice the length it was supposed to be.

Some students also presented content that was already covered in class, thereby not presenting anything new, or presented material that promoted astrology as a science or in 2012, the Maya Doomsday hype. To attempt to mitigate this, if such a topic was submitted, I would work with the students to make sure they presented aspects of that topic that would not be covered in class. For example, students have desired to present "extreme life" on Mars or Europa, which is covered in the class material in the section on Astrobiology. I suggested that they consider the extreme environments of the exoplanets and tie-in our understanding of Earth-based extreme life (as we know it) to that topic. That way, the material is new for everyone. In the case of astrology or the Maya Doomsday scenario, I encouraged the students to present scientific counterarguments along with the discussion of their topic. In the case of students creating a video about astrology, that usually works well, and discussions of night-sky motions and constellations enhance the discussion. However, in the case of the Maya Doomsday scenario, the students did not heed my advice at all. The video was full of doomsday propaganda and myth, thus scoring low on the "science" and "scientific accuracy" components of the rubrics. The class loved the video, though, and chose it as the "People's Choice" winner that semester. (Many of the award winning videos can be found at http://campus.albion.edu/nzellner/teaching/ and others can be found by searching youtube.com for "Albion College Astronomy".)

Sources also need to be credible, and the instructions explicitly state that three sources ("*not* wikipedia.org as the primary source") need to be used and included in the reference list. Oftentimes, students use more than three sources, especially if the video consists mostly of images. Finally, font and sound sometimes do not translate well to the video, and especially in the case of sound, poor quality can detract from enjoying the video. After a few instances of this, I included in the instructions a statement to the effect that students should "… pay attention to the distance between the microphone and the speaker and double-check the volume of your video on more than one computer."

Finally, no matter how many reminders there are, students tend to procrastinate, and procrastination can lead to a low-quality video. Working in pairs usually allows students to divide responsibilities and overcome difficulties, and a little bit of "peer pressure" can keep students on schedule. With more than one set of eyes and hands doing the work, it is also easier to make a good recording, be creative with the script, and figure out how to edit and format the video. Challenges for students working in pairs, however, include settling on (then changing) ideas and finding a time to work together. With careful planning and reminders, however, the difficulties are usually minimized.

**Attendance**

Class viewing usually occurs at the mid-point in the semester, and it is not unusual for attendance to taper off. Most of the time, though, students are eager to watch their classmates' videos, and incentives (i.e., popcorn, fruit, other goodies – we're watching movies after all!) will motivate student attendance in class. In the case where a student does not attend and does not provide a reasonable excuse, a 10% deduction is applied to their final video grade. In the case where a student does provide a reasonable excuse, that student is allowed to watch the videos on his/her/their own and provide feedback to me. By being flexible, students are more apt to see this as a positive experience. In fact, when students are not in class or conveniently excuse themselves when their video is about to be shown, they have been good-naturedly chided by their classmates. In a class where the average size is 26, this reaction may be an exception, however.

**Assessment**

After several semesters of student videos, assessment was conducted to see how the video project enhanced student learning. Qualitative assessment encourages the students to think about the content of the video and to reflect on their actions to create it (e.g., French and Burrows, 2017); student comments (above) were collected via an in-class activity. Thus, the qualitative learning outcomes include knowledge, synthesis, and evaluation. Since almost all of these students will not become professional astronomers, it was important to me to see how working in groups and using the Internet for research affected their view of "life-long learning" related to Astronomy. Intriguingly, in post-video assessments, most students preferred to work in groups and conducted Internet searches for Astronomy-related topics (Table 3). Working is groups is supported by various studies focused on peer-peer instruction and group interactions (e.g., Brame and Biel, 2015; Johnson et al. 2008), while an increased tendency to conduct Internet searches for Astronomy-related topics supports the finding of Wittman (2009) who suggested that Astronomy classes can improve the attitude about science (in general) of non-science students.

**Table 3.** Results of pre- and post-video assessment questions, from 90 students in six semesters.

|  | Pre-Video | Post-Video | % Change |
|---|---|---|---|
| **Work Style** |  |  |  |
| Group | 32 | 37 | 15.6 |
| Alone | 47 | 43 | -8.5 |
| No Preference | 11 | 10 | -9.1 |
|  |  |  |  |
| **Web Searches** |  |  |  |
| Yes | 40 | 63 | 57.5 |
| No | 26 | 13 | -50.0 |
| No Preference | 24 | 14 | -41.7 |

## CONCLUSION

Astronomy is one of the most popular science topics and often stimulates interest in the public's perception of science in general (e.g., Heck and Madsen, 2013; Percy 2006; Fraknoi 1998). Thus, at a time when improved science communication is reported to be important (Farahi et al. 2018; Kenney et al. 2016; National Research Council, 2010; Burns et al. 2003; Fraknoi 1998), these students are becoming equipped – with knowledge and skills – to make valuable contributions to society beyond the classroom. Via the video project at Albion College, students can showcase talents and knowledge that may not be reflected in answers on exams. They additionally develop skills that allow them to create, edit, and publish a video; to work collaboratively with peers; and to learn some science during the process.

As Maria Mitchell (1818-189) said, "We especially need imagination in science. It is not all mathematics, nor all logic, but is somewhat beauty and poetry." The ultimate goal of the project – and perhaps of introductory Astronomy courses at all institutions in general – is for students to learn how to appreciate the wonder of the night sky … and how we know what we know. And by producing a video that encourages students to think about space in ways that are creative and beautiful, they do.

## ACKNOWLEDGEMENTS

Conference travel to present this work at the 2016 National Astronomy Teaching Summit was supported by a grant from the Hewlett-Mellon Fund for Faculty Development at Albion College, Albion. MI. NEBZ thanks student Victoria Della Pia, who sorted and classified the original lists of videos, and Albion College's Foundation for Research, Scholarship, and Creative Activity which provided funding for her; the Great Lakes College Association for funding to travel to the Global Liberal Arts Alliance science workshop in Athens, Greece in 2017, where details about assessment and learning outcomes were established; and Jocelyn McWhirter, Director of Albion College's Center for Teaching and Learning, for conversations about this project. Most of all, NEBZ thanks the creative students in 18 semesters of Phys 105. Video day is still her favorite day!

**Appendix 1**

Video Killed the Writing Assignment
(instructions)

Nicolle Zellner[1,*]
[1*] Department of Physics, Albion College, Albion, MI, USA 49224

nzellner@albion.edu

**Astronomy Videos!**
**Due on X/XX at 11:59 pm with viewing in class on X/XX (and X/XX).**

**Rationale:**
So, you know how you're in college and you have to write a lot of papers? Not today! I've decided to let you put your *liberal arts education* to good use! Your assignment is to create a video of a famous astronomer, astronomical discovery, or telescope observatory. You will work *together* in groups of two to make a 5-6-minute digital video. Your task is to be creative yet still include accurate scientific content. The video judged best by the class will receive a special prize – NASA "swag" – or whatever is behind "Door #2"…

Some ideas for you to consider include[1],[2]:

- ■ documenting how a great discovery was made;
- ■ chronicling the life of a great astronomer;
- ■ recording the moment something was discovered at an observatory; or
- ■ anything else that you think is appropriate and approved by me!

**Student Learning Outcomes:**
By the end of this activity, you will have conducted an in-depth investigation of a topic of your choice and become the foremost class expert on that topic (knowledge). You will also have learned to use technology to create an entertaining and educational video (synthesis). Finally, you will constructively critique your classmates' videos (evaluation).

**Instructions:**
A title page and a page of references (at least 3, *not* wikipedia.org as a primary source) should be included *in* the video. Additionally, a hard copy of the references should be turned in to me. Any website listed on the hard copy needs to have the access date listed, as well as a brief description.

Your video should be between around 5:30 ± 30 seconds. At the 6-minute mark, the video will be turned off and points will be lost. Additionally, any portion of the video over 6 minutes will not be considered. You should plan your recording sessions carefully in order to avoid editing. However, if you want to edit (and know how and have the capability to do so), that's fine with me. Information Technology (IT) is prepared to help you – just know that it can be time-consuming. **You will post your video on youtube.com, tagged with "YOUR COLLEGE". Alternatively, you can post it on the CAMPUS PUBLIC DRIVE in the CLASS folder or on the Google drive. Finally, you will send me an email letting me know you have done so; this email needs to include the title of your video and, if on youtube.com or the Google drive, the link.** You must make sure that your video runs on the equipment in Norris 102. Failure to do so, and thus causing a delay in the viewing session, will result in a reduced grade.

---

[1] Each group should have its own unique topic. There will be no duplicate topics in any one class.
[2] Check out http://www.storybehindthescience.org/astronomy.html for other ideas.

Content and accuracy are important, as is CREATIVITY. Your video will be evaluated for both its scientific content and its entertainment value. Have FUN! Each person in the group should collaborate equally to make a great little video!

**Equipment:**

If you want to use your own equipment, **YOU** are responsible for making sure we can see it and hear it in Norris 102.

If you are going to use equipment from IT, please be aware that you will need to reserve the equipment ahead of time by filling out the online form found on the IT website under "Sign-out equipment". It asks for a faculty sponsor name and email where you will put my name and email. When I receive the email, I will send my authorization to them and the equipment will be reserved.

Please note that this may take some time, so please plan accordingly.

In the past, there have been some issues with volume. Please pay attention to the distance between the microphone and the speaker and double-check the volume of your video on more than one computer. If we can't hear it, we can't enjoy the video or appreciate all the hard work you've done!

Finally, please note that the <span style="color:red">public drive</span> isn't appropriate for the final project files because that space is volatile. You can copy a video there before presentation time if necessary, but you shouldn't expect it to be there indefinitely.

**Evaluations:**

Each video will be evaluated by your classmates and by me with the following criteria:

> **Content:** Did the images, animations, or other illustrations enhance the video? Was the organization appropriate? Was the audio adequate? Was the organization appropriate?

> **Science:** Could you understand the science? Was it at the appropriate level?

> **Delivery:** How appealing was the presentation? Did it entertain you?

I will additionally evaluate the credibility of the references, the scientific accuracy of the video, and how the video adhered to the requirements.

> **Content:** Were the sources credible and appropriate?

> **Science:** Was the science appropriate? What the science accurate?

> **Adherence to Requirements:** Did the video stay within the time limits? Did the video contain a title slide? Was a bibliography included in the video? Was a reference page handed to me? Did you delay the viewing session at all because you hadn't tested your video ahead of time?

The average of all of your classmates' evaluations will count as 33% of your grade; my evaluation will make up the other 67%. If you do not participate in evaluating your classmates' videos and/or if you do not show up to watch the videos, 10% will be deducted from your grade.

**All videos will be posted on youtube.com, the Google drive, or on the CAMPUS PUBLIC DRIVE in the CLASS folder on X/XX (appropriate day of the week, prior to the viewing day). You will send me an email telling me this, with the title and/or link, no later than 11:59 pm on X/XX. We will then view them in class on X/XX (and X/XX). Absolutely no late videos will be accepted. Class attendance on viewing day is REQUIRED.**